\documentclass[twocolumn,showpacs,preprintnumbers,amsmath,amssymb,aps,prb]{revtex4}
\usepackage{graphicx}

\usepackage{dcolumn}
\usepackage{bm}
\usepackage{mathrsfs}
\usepackage{epsfig}
\usepackage{graphicx}
\usepackage{amsmath}
\usepackage{amsfonts}
\usepackage{amssymb}
\usepackage{mathrsfs}
\usepackage{multirow}

\begin{document}
\title{Vortex states in mesoscopic superconducting squares: 
Formation of vortex shells}

\author{H.J.~Zhao}
\affiliation{
Department of Physics, University of Antwerpen, Groenenborgerlaan 171, B-2020 Antwerpen,
Belgium
}
\author{V.R.~Misko}
\affiliation{
Department of Physics, University of Antwerpen, Groenenborgerlaan 171, B-2020 Antwerpen,
Belgium
}
\author{F.M.~Peeters}
\affiliation{
Department of Physics, University of Antwerpen, Groenenborgerlaan 171, B-2020 Antwerpen,
Belgium
}
\author{V.~Oboznov}
\affiliation{
Institute of Solid State Physics, Russian Academy of Sciences, Chernogolovka 142432, Russia
}
\author{S.V.~Dubonos}
\affiliation{
Institute of Solid State Physics, Russian Academy of Sciences, Chernogolovka 142432, Russia
}
\author{I.V.~Grigorieva}
\affiliation{
School of Physics and Astronomy, University of Manchester, Manchester M13 9PL, UK
}

\date{\today}

\begin{abstract}
We analyze theoretically and experimentally vortex configurations in mesoscopic
superconducting squares.
Our theoretical approach is based on the analytical solution of the London equation
using Green's-function method.
The potential-energy landscape found for each vortex configuration is then used
in Langevin-type molecular-dynamics simulations to obtain stable vortex
configurations.
Metastable states and transitions between them and the ground state are analyzed.
We present our results of the first direct visualization of vortex patterns in
$\mu$m-sized Nb squares, using the Bitter decoration technique.
We show that the filling rules for vortices in squares with increasing applied
magnetic field can be formulated, although in a different manner than in disks,
in terms of formation of vortex ``shells''.
\end{abstract}

\maketitle

\section{Introduction}

The growing interest in studying vortex matter in mesoscopic and
nano-patterned superconductors is closely related to recent
progress in nano-fabrication and perspectives of their use in
nano-devices manipulating single flux quanta.
As distinct from bulk superconductors, vortex states in nano- and
mesoscopic samples are determined by the interplay between the
intervortex interaction (which is modified due to the presence of
boundaries) {\it and} the confinement. In general, the shape of a
mesoscopic sample is incommensurate with the triangular Abrikosov
lattice, and as a consequence, the resulting vortex patterns display 
strong features of the sample shape and may differ strongly from 
a triangular lattice. 
Strong finite size effects in conjunction with strong {\it shape}
effects determine the vortex configurations.
For example, in mesoscopic disks vortices, as shown theoretically
\cite{geim_nat97,deo_prl97,sch_prb98,sch_prl99,baelus_prb01,baelus_prb04,slava_disk}
and experimentally \cite{irina06}, form circular symmetric shells
(similar to two-dimensional (2D) system of charged classical particles \cite{bedanov}).
Moreover, due to strong confinement effects in small disks vortices can even 
merge into a giant vortex (GV), i.e., a single vortex containing more than one 
flux quantum \cite{sch_prl99}, as was recently confirmed experimentally \cite{kanda}. 
Furthermore, it was recently demonstrated \cite{irina07} that vortices can merge 
into a cluster or 
a GV in $\mu$m-sized mesoscopic niobium disks which is induced by strong disorder 
in combination with rather {\it weak} confinement, while neither of these effects 
alone would lead to a GV/cluster formation. 
Similarly, shape- and symmetry-induced vortex patterns can be formed in mesoscopic
superconducting triangles
\cite{vvm_vavnat,slava_vav,baelus_prb02},
squares
\cite{baelus_prb02,melnikov,kabanov,geurts2007}, 
or, in general, in symmetric polygons \cite{chi_phc02,baelus_prb02}. 
However, unlike disks where the vortex patterns result from the interplay
between the discrete symmetry of the (triangular) vortex lattice and the
cylindrical ($C_{\infty}$) symmetry of the disk, mesoscopic polygons have
discrete symmetry that can coincide (triangles, $C_{3}$ symmetry) or
include as a subgroup (e.g., hexagons with $C_{6}$ symmetry) the
symmetry of the vortex lattice.
In such cases highly stable vortex configurations are possible for some
values of magnetic field (providing commensurate numbers of vortices)
because the vortex-vortex interaction is {\it enhanced} by the effect
of boundaries.
Strikingly, strong boundary effects can even lead to symmetry-induced
vortex states with antivortices
\cite{vvm_vavnat,slava_vav,geurts2007}
(i.e., the symmetry of the vortex configuration with antivortices can be
restored by the generation of a vortex-antivortex pair). 

In contrast to $C_{3n}$-symmetric (where $n$ is an integer)
polygons, squares are {\it incommensurate} with triangular vortex
lattice for {\it any} applied magnetic field. The vortex-vortex
interaction and the effect of boundaries are always {\it competing} 
in mesoscopic squares. Resulting from this interplay: i) the ground 
state of the vortex system always involves nonzero elastic energy
and, as a consequence, ii) there are metastable states with
energies close to the ground state (or, in principle, the ground
state even could be degenerate).
Early studies on vortices in mesoscopic squares were either limited to very small
samples with characteristic sizes of the order of
$\xi$ (where $\xi$ is the coherence length) which were able to accommodate only
few vortices \cite{baelus_prb02}, or they focused on the possibility of generation
and stability of vortex-antivortex patterns in squares
\cite{melnikov,kabanov,geurts2007}.
Here we present a systematic theoretical analysis of vortex
configurations in mesoscopic squares and their first direct
observation in $\mu$m-sized niobium squares using the Bitter
decoration technique. To study the formation of vortex patterns
and transitions between the ground and metastable states, we
analytically solve the London equation using the Green's function
method, and perform molecular-dynamics simulations. 
To obtain the stable vortex configurations, we 
analyze the filling of squares by vortices with increasing applied 
magnetic field and the formation of vortex ``shells'', similarly 
to those observed in disks. 

The paper is organized as follows. 
The theoretical formalism and the solution of the London equation 
using the Green's function method, for a system of $L$ vortices in 
a rectangle sample, are described in Sec.~II. 
In Sec.~III, we discuss the evolution of vortex configurations with 
magnetic field calculated using the solution of the London equation 
found in Sec.~II and the molecular-dynamics simulations (Sec.~III.A). 
We formulate the filling rules and discuss the formation 
of vortex shells in mesoscopic superconducting squares in Sec.~III.B. 
Metastable states and the transitions between them and the ground state 
are analyzed in Sec.~III.C. 
In Sec.~IV, we present the results of our direct experimental observations  
of vortex patterns in niobium squares using the Bitter decoration technique, 
and compare the calculated patterns with the experimentally measured 
vortex configurations. 
The conclusions are given in Sec.~V.

\section{Theory: The London approach}

\begin{figure}
  \includegraphics[width=0.8\columnwidth]{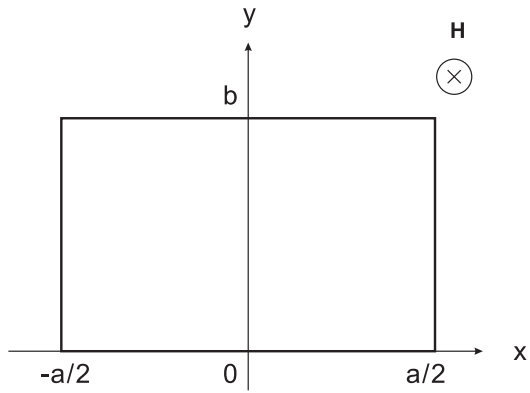}\\
  \caption{The cross-section of a rectangular superconductor
  with sides $a$ and $b$. The external magnetic field $\textbf{H}$
  is applied along the $z$-axis, and its value is assumed to be constant
outside the sample. }
\label{fig-coordinate}
\end{figure}

We consider a strong type-II superconductor (i.e., characterized 
by the Ginzburg-Landau parameter $\kappa=\lambda/\xi\gg1$, where 
$\lambda$ is the London penetration depth and $\xi$ is the 
coherence length) with  rectangular cross section in the $x$-$y$
plane 
and thickness $d$ in the $z$-direction. 
Note that the London approach is applicable also for {\it weak} 
type II superconductors in case of thin-film samples with 
thickness $d \ll \lambda $ where the penetration depth is 
modified: $\lambda \to$ $\Lambda = \lambda / d^{2}$, or in case of 
low vortex densities in rather {\it large} mesoscopic samples 
(i.e., with the lateral dimensions $a$, $a \gtrsim \Lambda$) 
where vortices are well separated and the order parameter is 
$|\Psi|^{2}=1$ everywhere except at the vortex cores. The latter 
case corresponds to our experiments with $\mu$m-sized niobium 
squares as described below. 
In our model the external magnetic field $\textbf{H}$ is applied normal to the $x$-$y$
plane, i.e., along the $z$-axis: $\textbf{h}=h\textbf{z}$.
We also assume that the vortex cores are straight lines along the
$z$-direction.
Then the local magnetic field can be found by solving the London equation:
\begin{equation}\label{Eq:london}
-\lambda^{2}\nabla^{2}h+h=\Phi_0 h
\sum_{i=1}^{L}\delta(\boldsymbol{r}-\boldsymbol{r}_i),
\end{equation}
where $\Phi_0$ is the flux quantum and $\{\textbf{r}_i=(x_i,y_i)$,
$i=1,\ldots,L\}$ are the positions of $L$ vortices. If we also 
neglect the distortion of the external magnetic field due to the
sample, i.e., assume that the value of the magnetic field outside
the sample near its boundary is equal to the applied field, then
the boundary conditions for the magnetic field are:
\begin{equation}
\label{eq-bf}
h(\pm a/2,y)=h(x,0)=h(x,b)=H.
\end{equation}
The geometry of the problem is shown in Fig. \ref{fig-coordinate}.
The Green's function method for solving the London equation 
Eq.~(\ref{Eq:london}) with the boundary conditions 
Eq.~(\ref{eq-bf}) was previously used by Sardella {\it et al.} 
\cite{edson}. However, they limited themselves to the special case 
where one of the sides of the rectangle is much larger than the 
other, i.e., {\it a stripe}. Such an approximation 
considerably simplifies the problem but the resulting solution 
missed the generality (the symmetry with respect to the 
permutation $x \to y$) and thus could not be used in our case of a 
square: $a=b$. We seek for a solution of Eq.~(\ref{Eq:london}) 
with the boundary conditions Eq.~(\ref{eq-bf}) which is valid for 
a rectangle with arbitrary aspect ratio $a/b$. 
The Green's function associating with the boundary problem defined by
Eqs.~(\ref{Eq:london}) and (\ref{eq-bf}) must satisfy the following equation:
\begin{equation}\label{eq:Green}
-\lambda^{2}\nabla^{2}G+G=\delta(x-x^\prime)\delta(y-y^\prime) ,
\end{equation}
and the boundary conditions:
\begin{equation}\label{eq:Gb1}
G(\pm a/2,y)=G(x,0)=G(x,b)=0 .
\end{equation}
Multiplying Eq. (\ref{Eq:london}) by $G$ and Eq. (\ref{eq:Green})
by $h$ and subtract one from another, we obtain
\begin{align}\label{eq:dr1}
-\lambda^{2}&(G\nabla^{2}h-h\nabla^{2}G)\nonumber\\
&=G\Phi_0\sum_{i=1}^{L}\delta(\boldsymbol{r}-\boldsymbol{r}_i)-h\delta(x-x^\prime)\delta(y-y^\prime).
\end{align}
Integrating Eq. (\ref{eq:dr1}) over the sample area, we arrive at
\begin{align}
-\lambda^{2}&\int_{-a/2}^{a/2}dx\int_{0}^{b}dy(G\nabla^{2}h-h\nabla^{2}G)\nonumber\\
=&\int_{-a/2}^{a/2}dx\int_{0}^{b}dy\big[G\Phi_0\sum_{i=1}^{L}\delta(\boldsymbol{r}-
\boldsymbol{r}_i)-\nonumber\\&h\delta(x-x^\prime)\delta(y-y^\prime)\big].\label{eq:dr2}
\end{align}
Further we use Gauss theorem,
\begin{align}
-&\lambda^{2}\int_{-a/2}^{a/2}dx\int_{0}^{b}dy(G\nabla^{2}h-h\nabla^{2}G)\nonumber\\
&=-\lambda^{2}\oint_{boundary}dl\big(G\frac{\partial h }{\partial
n}-h\frac{\partial G}{\partial n}\big)\nonumber ,
\end{align}
where $\partial / \partial n$ is the derivative in the normal
direction to the boundary,
and the boundary conditions Eqs. (\ref{eq:Gb1}) and (\ref{eq-bf}), and we find the expression
for the magnetic field:
\begin{align}
h(x^\prime,y^\prime)=&H\bigg[1-\int_{-a/2}^{a/2}dx\int_{0}^{b}dy
G(x,y,x^\prime,y^\prime)\bigg]\nonumber\\
&+\Phi_0\sum_{i=1}^{L}G(x_i,y_i,x^\prime,y^\prime). \label{eq:hgf}
\end{align}
Therefore, the problem of finding the solution for the local magnetic field is reduced to the
determination of the Green's function $G(x,y,x^\prime,y^\prime)$.
In order to find a solution to Eq. (\ref{eq:Green}) with the
boundary condition Eq. (\ref{eq:Gb1}), we expand the Green's
function in a Fourier series,
\begin{equation}
\label{eq:dg1}
G(x,y,x^\prime,y^\prime)=\frac{2}{b}\sum_{m=1}^{\infty}\sin(\frac{m\pi
y^\prime}{b})\sin(\frac{m\pi y}{b})g_m(x,x^\prime).
\end{equation}
Note that the boundary conditions Eq. (\ref{eq:Gb1}) are satisfied at $y=0,b$.
Further we substitute this expansion into Eq. (\ref{eq:Green}) and obtain
\begin{eqnarray}\label{eq:dg2}
&&-\lambda^2\frac{2}{b}\sum_{m=1}^{\infty}\bigg[\frac{\partial^2
g_m (x,x^\prime)}{\partial x^2 }\sin(\frac{m\pi
y^\prime}{b})\sin(\frac{m\pi y}{b})\nonumber\\
&&-(\frac{m\pi}{b})^2 g_m(x,x^\prime)\sin(\frac{m\pi
y^\prime}{b})\sin(\frac{m\pi
y}{b})\nonumber\\
&&+\sin(\frac{m\pi y^\prime}{b})\sin(\frac{m\pi
y}{b})g_m(x,x^\prime)\bigg]\nonumber\\
&&
=\delta(x-x^\prime)\frac{2}{b}\sum_{m=1}^{\infty}\sin(\frac{m\pi
y^\prime}{b})\sin(\frac{m\pi y}{b}),
\end{eqnarray}
where we used the following $\delta$-function representation
\begin{eqnarray}
\delta(y-y^\prime)=\frac{2}{b}\sum_{m=1}^{\infty}\sin(\frac{m\pi
y^\prime}{b})\sin(\frac{m\pi y}{b})
\nonumber
\end{eqnarray}
since $ \Big\{\sqrt{\frac{2}{b}}\sin(\frac{m\pi y}{b}), \
m=1,2,3\ldots\Big\}\nonumber$
 forms a complete set of orthonormal
functions. As a result, we obtain the following equation for the
Fourier-transform of the Green's function $g_m (x,x^\prime)$,
\begin{eqnarray}\label{eq:gm}
-\lambda^2\frac{\partial^2 g_m (x,x^\prime)}{\partial x^2
}+\alpha_{m}^{2} g_m (x,x^\prime)=\delta(x-x^\prime), 
\end{eqnarray}
where 
\begin{equation}
 \alpha_m=\bigg[1+\lambda^2\big(\frac{m\pi}{b}\big)^2\bigg]^{1/2}\label{eq:am}.
\end{equation}
The functions $g_m(x,x^\prime)$ must satisfy the boundary conditions $ g_m (\pm a/2,x^\prime)=0$.
In order to solve Eq. (\ref{eq:gm}), we first take its Fourier transform,
\begin{equation}
-\lambda^2(i\omega)^2F(\omega)+\alpha_{m}^{2}
F(\omega)=\frac{1}{2\pi}e^{-i\omega x^\prime},\nonumber
\end{equation}
where
\begin{equation}
F(\omega)=\frac{e^{-i\omega
x^\prime}}{2\pi(\lambda^2\omega^2+\alpha_{m}^{2})},
\nonumber
\end{equation}
from which we obtain a particular solution to Eq. (\ref{eq:gm})
\begin{align}
&g_m|_{a\rightarrow\infty}=\frac{1}{2\alpha_m\lambda}e^{-\alpha_m|x-x^\prime|/\lambda}\nonumber\\
&=\frac{1}{2\alpha_m\lambda}\big[\cosh(\alpha_m
(x-x^\prime)/\lambda)-\sinh(\alpha_m|x-x^\prime|\lambda)\big].
\nonumber
\end{align}
The general solution of Eq. (\ref{eq:gm}) reads as:
\begin{align}
g_m=&\frac{1}{2\alpha_m\lambda}\big[\cosh(\alpha_m
(x-x^\prime)/\lambda)-\sinh(\alpha_m|x-x^\prime|\lambda)\big]\nonumber\\
&+A(x^\prime)\sinh(\alpha_m
x/\lambda)+B(x^\prime)\cosh(\alpha_m x/\lambda)\nonumber\\
=&\frac{1}{2\alpha_m\lambda}\big[-\sinh(\alpha_m|x-x^\prime|\lambda)+C(x^\prime)\sinh(\alpha_m
x/\lambda)\nonumber\\
&+D(x^\prime)\cosh(\alpha_m x/\lambda)\big].
\nonumber
\end{align}
Using the boundary conditions Eq. (\ref{eq:Gb1}) we find the
coefficients $C(x^\prime)$ and $D(x^\prime)$:
\begin{align}
&C(x^\prime)=-\coth(\alpha_m a/2\lambda)\sinh(x^\prime);\nonumber\\
&D(x^\prime)=\tanh(\alpha_m a/2\lambda)\cosh(x^\prime).\nonumber
\end{align}
Then the solution for $g_m(x,x^\prime)$ is given by
\begin{align}\label{eq:gmf}
g_m(x,x^\prime)=&\frac{1}{2\lambda\alpha_m \sinh(\alpha_m
a/\lambda)}\nonumber\\
&\times\bigg\{\cosh\big[\alpha_m(|x-x^\prime|-a)/\lambda\big]\nonumber\\
&-\cosh\big[\alpha_m(x+x^\prime)/\lambda\big]\bigg\}.
\end{align}
Inserting this result into Eq. (\ref{eq:dg1}), we obtain the 
expression for the Green's function: 
\begin{align}
G(x,&y,x^\prime,y^\prime)=\frac{2}{b}\sum_{m=1}^{\infty}\sin(\frac{m\pi
y^\prime}{b})\sin(\frac{m\pi
y}{b})\nonumber\\
&\times\frac{1}{2\lambda\alpha_m \sinh(\alpha_m
a/\lambda)}\bigg\{\cosh\big[\alpha_m(|x-x^\prime|-a)/\lambda\big]\nonumber\\&
-\cosh\big[\alpha_m(x+x^\prime)/\lambda\big]\bigg\}.
\end{align}
From it we obtain an expression for the local magnetic field:
\begin{align}
h(x,y)=&\Phi_0\sum_{i=1}^{L}G(x_i,y_i,x,y)+H\bigg\{\frac{\cosh[(y-b/2)/\lambda]}{\cosh(b/2\lambda)}\nonumber\\
&+\frac{4}{b}\sum_{m=0}^{\infty}\frac{b}{\alpha_{2m+1}^{2}(2m+1)\pi}\sin\bigg[\frac{(2m+1)\pi
y}{b}\bigg]\nonumber\\
&\times\frac{\cosh(\alpha_{2m+1}x/\lambda)}{\cosh(\alpha_{2m+1}a/2\lambda)}\bigg\}.
\end{align}
Note that this solution is valid for a rectangle with arbitrary 
aspect ratio $a/b$ and is a generalization of the earlier result 
presented in Ref.~\onlinecite{edson}. 
\begin{widetext}
Using the obtained solution or the London equation for the local
distribution of the magnetic field $h(x,y)$, we obtain the Gibbs
free energy per unit length of an arbitrary vortex configuration:
\begin{align}
\mathscr{G}=&\sum_{i=1}^L\bigg(\epsilon_i^{shield}+\sum_{j=1}^L\epsilon_{ij}^v\bigg)
+\epsilon^{core}+\epsilon^{field}\nonumber\\
=&\frac{\Phi_0 H}{4\pi A}\sum_{i=1}^{L}\bigg\{\frac{\cosh[(y_i-b/2)/\lambda]}{\cosh(b/2\lambda)}
+ \frac{4}{b}\sum_{m=0}^{\infty}\alpha_{2m+1}^{-2}\frac{b}{(2m+1)\pi}\sin \big [\frac{(2m+1)\pi y_i}{b}\big]\frac{\cosh(\alpha_{2m+1}x_i/\lambda)}{\cosh(\alpha_{2m+1}a/2\lambda)}\bigg\}\nonumber\\
 +&\frac{\Phi_0^2}{8\pi A}\sum_{i=1}^L\sum_{j=1}^{L}G(x_i,y_i,x_j,y_j)-\frac{H^2}{8\pi}\bigg\{\frac{\tanh(b/2\lambda)}{b/2\lambda}
-\frac{8}{\pi^2}\sum_{m=0}^{\infty}\frac{\tanh(\alpha_{2m+1}a/2\lambda)}{[(2m+1)\alpha_{2m+1}]^2(\alpha_{2m+1}a/2\lambda)}\bigg\}\nonumber\\
-&L\frac{\Phi_0 H}{4\pi A}.
\label{eq:freef}
\end{align}
\end{widetext}
Here, $A = a \times b$ is the area of the rectangle. 
The last two terms are the energies associated with the external
magnetic field and the vortex cores, respectively. The Green's
function in the first term describes the interaction between
vortices and also the interaction between vortices and their
images which are situated outside the sample. The second term
represents the interaction between the $i$th vortex and the
shielding currents. Note that in Ref.~\onlinecite{edson}, the
authors limited their consideration to the case of a thin film
such that $(\pi\lambda/b)^2\gg 1$ and the term ``$1$" in 
Eq.~(\ref{eq:am}) can be neglected. 
%
%
The London theory has a singularity for the interaction between a
vortex and its own image (self-interaction). We notice that when
$i=j$ the Green's function does not converge. To avoid the
divergency, we apply a cutoff procedure (see, e.g.,
Refs~\onlinecite{Abrikosov,Buzdin,Ketterson}), which means a
replacement of $|\textbf{r}_i-\textbf{r}_j|$ by $a\xi$ for $i=j$.
It was shown in Ref.~\onlinecite{Cabral} that the 
results of the London theory agree with those of the 
Ginzburg-Landau theory, the vortex size should be chosen as 
$\sqrt2\xi$, and therefore we take $a=\sqrt2$. 
%
%
The confinement energy is given by
$\epsilon_c=\epsilon_i^{shield}+\epsilon_{ii}$.

\begin{figure}
  \includegraphics[width=0.65\columnwidth]{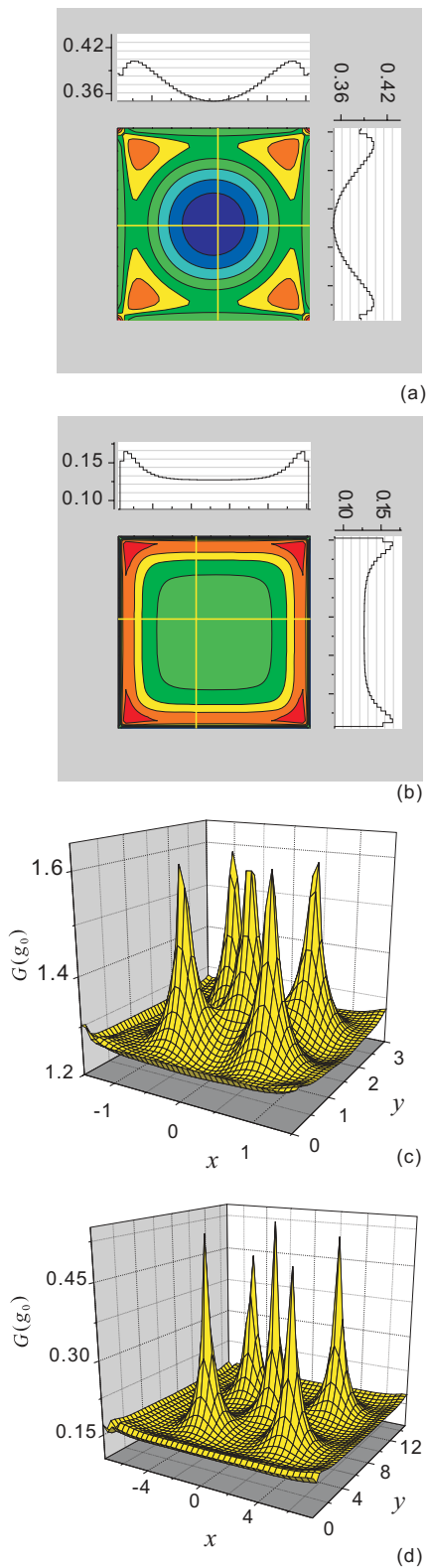}
  \caption{(Color online)
The profiles of the confinement energy 
$\epsilon_c=\epsilon_i^{shield}+\epsilon_{ii}$ 
(measured in units of
$g_0=\Phi_0^2/8\pi A \cdot 1/\lambda^2$, where $A$ is the area of
the sample) for mesoscopic superconducting squares with size
$a=3\lambda$ (a) and $15\lambda$ (b). The Gibbs free energy 
distributions for squares with $a=3\lambda$ (c) and $15 \lambda$ (d) for
the vortex state with $L=5$.
}
\label{fig:confinement}
\end{figure}

In Figs~\ref{fig:confinement}(a) and (b), we plot the distribution of the confinement energy
for mesoscopic squares with $a=3\lambda$ and $a=15\lambda$, correspondingly.
In the mesoscopic square with $a=3\lambda$, Fig.~\ref{fig:confinement}(a),
the screening current extends inside the square and interacts with all the vortices.
But in the large mesoscopic square (we call it ``macroscopic'') with $a=15\lambda$, only the vortices
which are close to the boundary feel the screening current.
In the mesoscopic square, vortices strongly overlap with each other (see Fig. \ref{fig:confinement}(c))
while in the macroscopic square, the interaction between vortices is rather weak and only the closest
neighbors are important (see Fig. \ref{fig:confinement}(d)). 
This difference between small (mesoscopic) and large (macroscopic) squares 
leads, in general, to the size-dependence of the vortex patterns in mesoscopic 
samples as it was recently demosntrated for disks 
(see Ref.~\onlinecite{slava_disk}). 

\section{The evolution of vortex patterns with magnetic field}

\subsection{Molecular-Dynamics simulations of vortex patterns}

Within the London approach, vortices can be treated as point-like 
``particles'', and it is convenient to employ Molecular Dynamics 
(MD) for studying the vortex motion driven by external forces 
(see, e.g., 
Refs.~\onlinecite{slava_disk},~\onlinecite{irina07},~\onlinecite{md01},
~\onlinecite{penrose}), 
similarly to a system of classical particles \cite{bedanov}. 
In the previous section we obtained the analytic expression for
the free energy of a system of $L$ vortices as a function of the
applied magnetic field, Eq.~(\ref{eq:freef}).
The force felt by the $i$th vortex can be obtained by taking the
derivative of the energy:
\begin{equation}
 F_i=-\nabla_i \mathscr{G}, 
\label{eq:fs}
\end{equation}
where $\nabla_i=\frac{\partial}{\partial
x_i}\textbf{e}_x+\frac{\partial}{\partial y_i}\textbf{e}_y$ is the
two dimensional derivative operator.

The overdamped equation of vortex motion can be presented in the form:
\begin{eqnarray}
\eta \textbf{v}_i=\textbf{F}_i=\sum_{j\neq
i}\textbf{F}_{ij}+\textbf{F}_{self}^{i}
+\textbf{F}_{M}^{i}+\textbf{F}_T^i .
\label{eq:ft}
\end{eqnarray}
where the first three terms are as follows: {\bf F}$_{ij}$ is the
force due to the repulsive vortex-vortex interaction of the $i$th
vortex with all other vortices, {\bf F}$_{self}^{i}$ is the
interaction force with the image, and {\bf F}$_{M}^{i}$ is the
force of interaction with the external magnetic field which enters
the sample through the boundaries; $\eta$ is the viscosity, which is
set here to unity.
Note that Eq.~(\ref{eq:fs}) contains these three terms (with the
free energy defined by Eq.~(\ref{eq:freef})), and in
Eq.~(\ref{eq:ft}) we added a thermal stochastic term
$\textbf{F}_T^i $ to simulate the process of annealing in the
experiment.
The thermal stochastic term should obey the following conditions: 
\begin{eqnarray}
\langle F_i^T(t)\rangle =0
\end{eqnarray}
and
\begin{eqnarray}
\langle F_i^T(t) F_i^T(t^\prime)\rangle =2\eta k_B
T\delta_{ij}\delta(t-t^\prime).
\end{eqnarray}
It is convenient to express the lengths in units of $\lambda$,
the fields in units of $H_{c2}$, the energies per unit length in units
of $g_0=\Phi_0^2/8\pi A \cdot 1/\lambda^2$, and the force per unit length
in units of $f_0=\Phi_0^2/8\pi A\cdot 1/\lambda^3$, where $A$ is the
sample's area.
In our calculations we use the value of the Ginzburg-Landau parameter
$\kappa=6$ taken from the experiment with Nb (see below).

In order to find the ground state vortex configurations in
squares, we perform stimulated annealing simulations by
numerically integrating the overdamped equations of motion
Eq.~(\ref{eq:ft}). The procedure is as follows. First we generate
a random vortex distribution and set a high value of temperature.
Then we gradually decrease the temperature to zero, i.e.,
simulating the annealing process in real experiments (see, e.g., 
Ref.~\onlinecite{harada}). To find the minimum energy 
configuration, we perform many simulation runs with random 
initial distributions and count the statistics of the appearance 
of different vortex configurations for each $L$. This procedure 
simulates \cite{slava_disk} the statistical analysis of 
experimental data with simultaneous measurements of vortex
configurations in arrays of {\it many} (up to 300) practically
identical samples. It was used in experiments with Nb disks in
Refs.~\onlinecite{irina06,irina07} and also in experiments with Nb
squares presented in this paper.

\subsection{Filling rules for vortices in squares with increasing
magnetic field: Formation of vortex shells}

\begin{figure}[tt]
\begin{center}
\includegraphics[width=0.95\columnwidth]{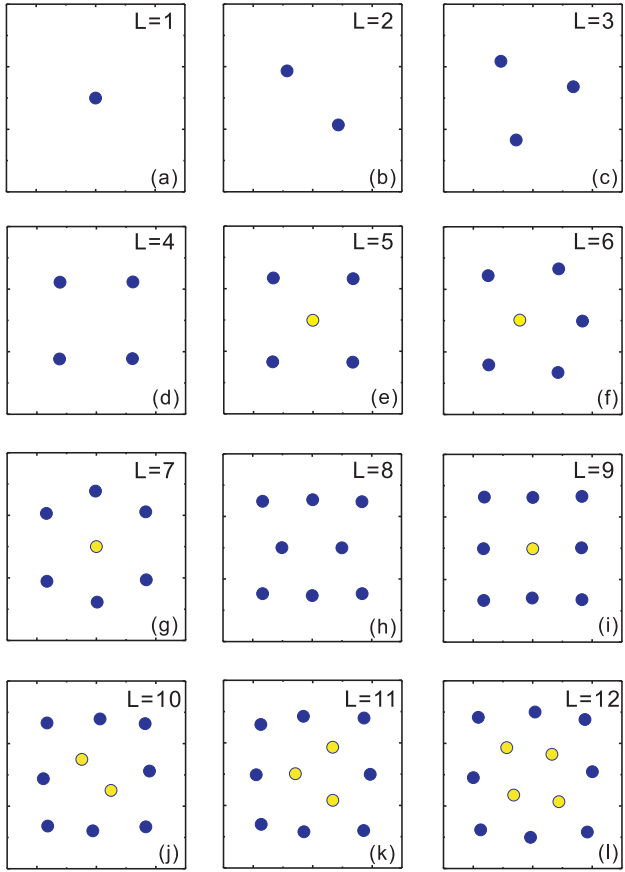}
\caption{(Color online)
The evolution of vortex configurations for the states with vorticity
increasing from $L=1$ to $12$, in a superconducting square with $a=3\lambda$
(the same results found for larger squares, e.g., with $a=15\lambda$). 
The vortices in the outer shell are shown by the blue (black) circles while the
inner-shell vortices are shown by the yellow (grey) circles.
The formation of the second shell starts when $L=5$. }
\label{fig:s1-6}
\end{center}
\end{figure}

\begin{figure}[t]
\includegraphics[width=0.95\columnwidth]{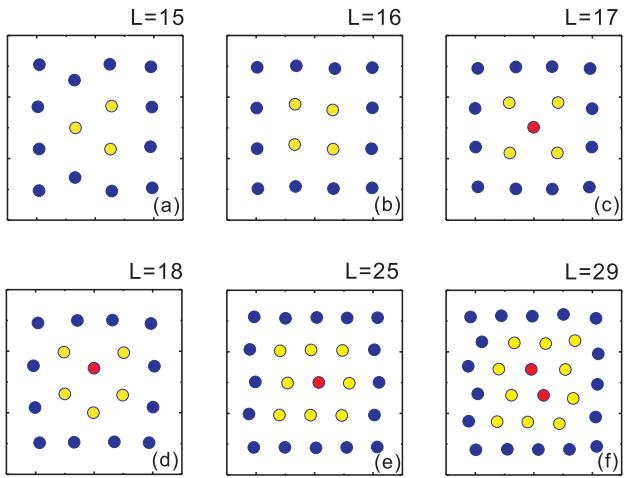}
\caption{(Color online)
The evolution of vortex configurations for 
$L=15$ to $18$ (a)-(d), and for $L=25$ (e) to $29$ (f), 
in a superconducting square with $a=3\lambda$. 
For vorticities $L=15$ to $18$ ((a)-(d)), the outermost shell formed
by 12 vortices is complete (commensurate with the square boundary),
and with increasing magnetic field vortices fill inner shells.
Note that when the inner shell also becomes complete ($L=16$, state (4,12) (b)),
the third shell starts to form for $L = 17$ (c).
For states with larger vorticities, e.g., $L=25$ (e) $L=29$ (f),
the vortex patterns are very close to a triangular lattice 
which is distorted near the boundary.} 
\label{fig:s29}
\end{figure}

The results for the vortex patterns for different vorticities $L$ 
are shown in Figs~\ref{fig:s1-6} and \ref{fig:s29}. 
With increasing applied magnetic field, vortex configurations 
evolve as follows: Starting from a Meissner state with no vortex, 
the first vortex appears in the center -- see Fig.~\ref{fig:s1-6}(a), 
for $L=2$ the two are located symmetrically on the diagonal -- see 
Fig.~\ref{fig:s1-6}(b). Further increase of the magnetic field 
leads to the formation of a triangular vortex pattern 
having a common symmetry axis with the 
square, which is the diagonal -- see Fig.~\ref{fig:s1-6}(c). 
For $L=4$ vortices arrange themselves in a perfect square, 
Fig.~\ref{fig:s1-6}(d), whose symmetry is commensurate with the 
sample and therefore it turns out that this is a highly stable 
vortex configuration\cite{baelus_prb01,Baelus}. 
Note that even in the bulk the gain 
in the elastic energy is very small during the transition from the 
triangular vortex lattice to the square one, and consequently, in 
the presence of a square boundary, it turns out that a square 
vortex lattice can be easily stabilized (for commensurate vortex 
numbers). 
For vorticity $L=5$, vortices tend to form either a pentagon, or a
square with one vortex in the center (see Fig.~\ref{fig:s1-6}(e),
the transition between this configuration and the pentagon-like
pattern will be discussed below). The additional vortex appears in
the center thus forming a second {\it shell } in a similar way as
in disks \cite{baelus_prb04,irina06,slava_disk}, but in the latter,
this occurred for a larger $L$-value ($L=6$). To distinguish 
different shells and indicate the number of vortices in each shell, 
we use the same notations as in 
Refs.~\onlinecite{baelus_prb04,irina06,slava_disk}. For example,
the pentagon-like configuration and the pattern with four vortices 
in the outershell and one vortex in the center are denoted as $(5)$ 
and $(1,4)$, respectively. (It is clear that vortex shells in squares
are not as well defined as in disks and sometimes it is a matter
of choice how to define them.)
Compared with disks, which have $C_\infty$ symmetry, the $C_4$
symmetry of squares induces a new element of symmetry in the
resulting vortex patterns. In other words, vortex patterns in
squares (tend to) acquire elements of the $C_4$ symmetry even if
they are not arranged in a perfect square lattice. For example,
the calculated vortex patterns share one ($L=6$,
Fig.~\ref{fig:s1-6}(f)) or two ($L=7$ and 8,
Figs.~\ref{fig:s1-6}(g) and (h), correspondingly) symmetry axes of
the square parallel to its side.
This tendency to share symmetry elements with the square boundary
remains also for larger vorticities as can be seen, e.g., in
Figs.~\ref{fig:s1-6}(j), (k), and (l) for vorticities $L = 10$,
11, and 12, respectively. For the commensurate number of vortices
$L=9$, a perfect symmetric square-lattice pattern is formed.

Using the concept of vortex shells, we analyzed the filling rules
for mesoscopic superconducting squares with increasing magnetic field.
To summarize these rules,
for $L=1$ to 4, vortices are arranged in a single ``shell''; the
second shell appears when $L=5$, and then vortices fill the shells 
as follows: 
As the vorticity $L$ increases from $L=5$ to 9, the new vortices
fill the outer shell. Then the number of vortices in the inner
shell starts to increase for $L \geq 9$ (see
Figs.~\ref{fig:s1-6}(j), (k), and (l)). This occurs because the
outer shell is formed by $8$ vortices (i.e., three per each side)
which turns out to be stable. Thus, the new vortices fill the
inner shell until $L=12$.
Then, again, the newly generated vortices start to fill the outermost
shell until $L=16$, when the number of vortices in the outermost shell
becomes $12$, which is also stable (i.e., commensurate with the
square boundary).
The formation of the third shell starts when the vorticity becomes
$L=17$ (note that for $L=17$ the vortices can arrange themselves either
in a two-shell configuration (5,12), or in a three-shell configuration
(1,4,12) which occurs to have a slightly lower energy, see analysis
below).
In a similar way, the filling of shells occurs for larger values of $L$ 
(e.g., for 3-, 4-shell patterns, etc.). 
As a general rule, the outermost shells containing $4N$ vortices, where
$N$ is an integer, are very stable.
With increasing the density of vortices, the average distance between
them decreases.
As a result, the interaction between vortices becomes more and more
important leading to the formation of the triangular-lattice phase
away from the boundary.
Therefore, the triangular lattice is recovered for large vorticities being 
distorted near the square boundaries.
Note that for large enough $L$ vortices do not form a square lattice 
even for commensurate vortex numbers (e.g., for $L=25$, 36, etc.) 
as it does for $L=4$, 9, and 16. 
Some expamples of two- and three-shell vortex patterns are shown 
in Fig.~\ref{fig:s29}.

\subsection{The ground state and metastable states}

\begin{figure}[t]
\begin{center}
\includegraphics[width=0.95\columnwidth]{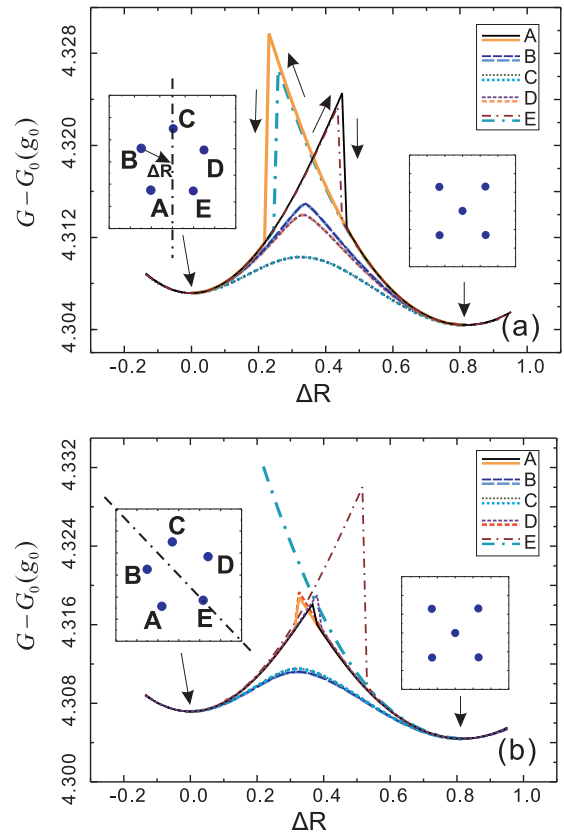}
\caption{(Color online)
The change of the free energy $(G-G_0)$ versus the displacement $R$
of one of the vortices in the initial pentagon-shaped configuration
from its initial position towards the center
(two different lines for each configuration correspond to 
increasing and decreasing $\Delta R$ as shown by the arrows in (a)).
$G_0$ is the free energy associated with external magnetic field 
and the vortex cores (term ``4'' in Eq.~(\ref{eq:freef})), which is independent 
of the positions of the vortices. 
The two stable states, the pentagon-like state (5) and the square
symmetric state (1,4), are shown in the insets.
The vortices are labeled by A, B, C, D and E.
Two different symmetry axes of the configuration (5) are shown by the
dash-dotted line in the insets of (a) and (b), respectively.
The side of the square is $a=3\lambda$.
In both cases, the configuration with one vortex in the center (1,4)
has a lower energy than the pentagon-like pattern (5).
Note that the curves for B and D (and for A and E) are slightly 
different due to the fact that the configuration (5) is not perfectly 
aligned with respect to the symmetry axes.}
\label{fig:metastable}
\end{center}
\end{figure}

The incommensurability of the square boundary 
with the triangular vortex lattice creates metastable vortex 
configurations. While in many cases metastable states are well 
separated in energy from the ground state, in some cases, namely, 
for borderline configurations having $n$ and $n+1$ shells, the 
lowest-energy metastable state can become almost indistinguishable 
from the ground state. 
In such cases, vortex states with very close energies can have
comparable probability to be realized experimentally. An example
of a such state is the case $L=5$.
The stable states for $L=5$ are shown in the insets of 
Fig.~\ref{fig:metastable}. In order to examine which one is more 
stable, we investigate the free energy as a function of the 
displacement of one of the vortices while we allow the other 
vortices to relax to their lowest-energy positions. 
We start with the pentagon-like configuration (5) (the left inset) 
and we change the position of this vortex moving it towards the 
center of the square and let the other vortices adjust their 
positions accordingly. At the end, we arrive at the 
square-symmetric state (1,4).
We plot the free energy of the system as a function of the displacement 
of this vortex from its equilibrium position, and we repeat this procedure 
for all the vortices A, B, C, D, and E (we always move only one vortex 
while all others relax to minimize the free energy). 
For any of the five vortices, this procedure leads to a barrier between 
the two states. 
We notice that there are two possible pentagon-like configurations
(5) which share different symmetry axes with the square, see
Figs.~\ref{fig:metastable}(a) and (b). The difference of their
free energy is less than $10^{-4}$.
In Fig.~\ref{fig:metastable}(a) we see that the motion of vortex C is
accompanied with the lowest energy barrier.
This is because vortices A, B, D and E are already close to their final
positions in state (1,4).
Moving vortices B or D lead to a higher energy barrier.
Finally, moving vortex A or E to the center is associated with the highest
barrier and passing over a saddle point (jump in $G-G_0$).
Then we move the central vortex of state (1,4) back to its initial
positions in state (5).
The highest-barrier transitions (i.e., curves A and E) show a hysteretic
behaviour which is an indication of metastable states.
\begin{figure}
\begin{center}
\includegraphics[width=0.95\columnwidth]{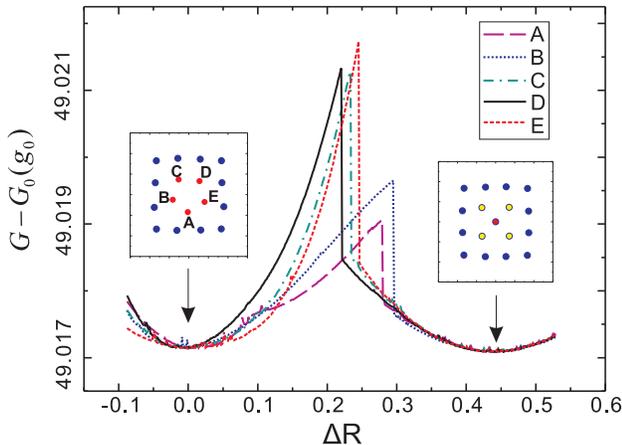}
\caption{(Color online)
The change of the free energy $(G-G_0)$ versus the displacement $R$ of one
of the vortex in the inner shell of the state (5,12) from its initial position
towards the center; $G_{0}$ is defined in the caption of Fig.~\ref{fig:metastable}.
The change in the free energy due to the movement of the vortices in the inner
shells (i.e., (5,12) $\to$ (1,4,12)) is damped by the movement of the vortices
in the outermost shell which act as a ``softer'' wall than the boundary
(in the case of the transition (5) $\to$ (1,4), see Fig.~\ref{fig:metastable}).
The movement of the vortices in the outmost shell causes more saddle points.
The two states, (5,12) and (1,4,12), have very close free energies. }
\label{fig:metastable17}
\end{center}
\end{figure}

In Fig.~\ref{fig:metastable}(b), 
we show the results of the calculation of the free energy as a function 
of displacement of a vortex, for a different modification of the state (5), 
i.e., when the vortex configuration has the symmetry axis coinciding with 
the diagonal of the square (cp. Fig.~\ref{fig:metastable}(a)). 
Note that these two configurations of state (5) have practically 
{\it the same} free energy and thus equal probability to appear 
in experiment. 
Moving vortex E, which is situated on the diagonal of the square 
(see the left inset in Fig.~\ref{fig:metastable}(b)), is accompanied by 
the highest energy barrier compared to moving other vortices. 
The reverse process (i.e., moving the central vortex to position E) leads
to a very high potential barrier, and the pentagon-like state cannot be
restored unless a random (thermal) force is added to break the symmetry.
Moving vortex B or C is accompanied by the lowest energy barrier.
State (1,4) has a lower free energy than state (5).
According to our calculations, it is the ground state for $L=5$.

Similar transitions are found between two- and three-shell vortex
configuration for $L=17$ (see Fig.~\ref{fig:metastable17}). Twelve
vortices form the outermost shell and the other five can form
either a one-shell or two-shell configurations similarly as state
$L=5$.
Again, we move one of the five vortices in the inner shell of the
state (5,12) to the center of the square. The analysis of the free
energy shows that the difference of the free energy between the
two states ($|\triangle G|\sim 10^{-5}$) is much smaller compared 
to the states for $L=5$ ($|\triangle G|\sim 10^{-3}$). The reason
for this is that for $L=17$, the twelve vortices in the outermost
shell can adjust themselves to lower the free energy, which create 
much ``softer'' walls for the five vortices in the inner shell 
than the sample boundary. 
Thus, the change of the free energy due to the movement of the vortices
in the inner shells can be more or less compensated by the movement of the
vortices in the outermost shell.

\section{Experimental observation of vortex configurations in mesoscopic Nb squares}

\begin{figure}
\begin{center}
\includegraphics[width=0.95\columnwidth]{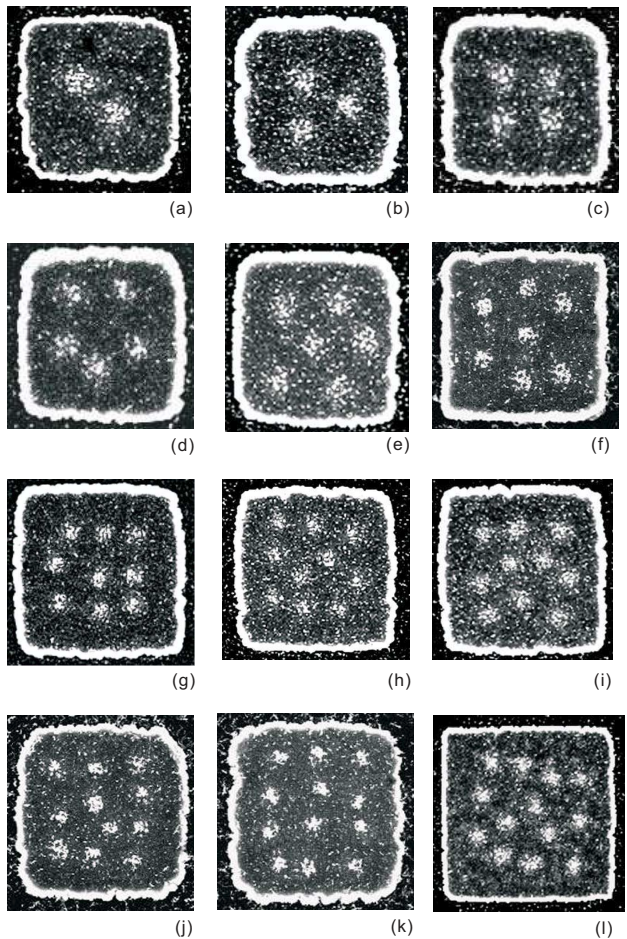}
\caption{
Scanning electron microscope (SEM) images of vortex configurations 
observed experimentally for vorticities $L = 2$ to 13. 
Vortex positions are indicated by clusters of small white (Fe) particles. 
(a) $L = 2$; sample size (side of the square) 
$a \approx 2.5$~$\mu$m, $H = 20$~Oe; 
(b) $L = 3$; $a \approx 2$~$\mu$m, $H = 35$~Oe; 
(c) $L = 4$; $a \approx 2.4$~$\mu$m, $H = 40$~Oe; 
(d) $L = 5$; $a \approx 2.4$~$\mu$m, $H = 40$~Oe; 
(e) $L = 6$; $a \approx 2.5$~$\mu$m, $H = 40$~Oe; 
(f) $L = 7$; $a \approx 2$~$\mu$m, $H = 60$~Oe; 
(g) $L = 9$; $a \approx 3.5$~$\mu$m, $H = 35$~Oe; 
(h) $L = 10$; $a \approx 3.5$~$\mu$m, $H = 35$~Oe; 
(i) $L = 10$; $a \approx 3.5$~$\mu$m, $H = 35$~Oe; 
(j) $L = 11$; $a \approx 2.5$~$\mu$m, $H = 60$~Oe; 
(k) $L = 12$; $a \approx 2.6$~$\mu$m, $H = 60$~Oe; 
(l) $L = 13$; $a \approx 5$~$\mu$m, $H = 20$~Oe.
}
\label{sqfig07}
\end{center}
\end{figure}

To visualise the corresponding vortex configurations experimentally 
we used the well-known Bitter decoration technique which is based on 
{\it in situ} evaporation of $10-20$~nm Fe particles that are attracted 
to regions of magnetic field created by individual vortices and thus 
allow their visualisation (details of the technique are described 
elsewhere \cite{irina94}). 
The mesoscopic samples for this study were made from a 150~nm thick 
Nb film deposited on a Si substrate using magnetron sputtering. 
The film's superconducting parameters were: transition temperature 
$T_{c} = 9.1$~K, magnetic field penetration depth $\lambda(0) \approx 90$~nm; 
coherence length $\xi(0) \approx 15$~nm; upper critical field 
$H_{c2}(0) \approx 1.5$~T. 
Using e-beam lithography and dry etching with an Ar ion beam through 
a 250~nm thick Al mask, the films were made into arrays of small square 
``dots'' of 4 different sizes, with the side of the square, $a$, varying 
from 1 to 5~$\mu$m. 
Each array typically contained 150 to 200 such dots. 
A whole array was decorated in each experiment, allowing us to obtain 
a snapshot of up to 100 vortex configurations in dots of the same shape 
and size, produced in identical conditions (same applied magnetic field 
$H$ and temperature $T$, same decoration conditions). 
It was therefore possible to simultaneously visualise vortex configurations 
for several different vorticities $L$ (in samples of different sizes) 
and also to gain enough statistics for quantitative analysis of the observed 
vortex states in terms of their stability, sensitivity to sample imperfections, 
and so on. 
Below we present the results obtained after field-cooling to $T \approx 1.8$~K 
in perpendicular external fields ranging from $H = 20$ to 60~Oe. 
We note that the above temperature (1.8~K) represents the starting temperature 
for the experiments. 
Thermal evaporation of Fe particles usually leads to a temporary increase 
in temperature of the decorated samples but the increase never exceeded 2~K 
in the present experiments, leaving the studied Nb dots in the low-temperature limit, 
$T < 0.5$~$T_{c}$. 

Fig.~7 shows examples of vortex configurations observed for vorticities $L = 2$ to 13. 
The images shown in Fig.~7 were obtained in several different experiments and on 
samples of different sizes (see figure caption). 
We note that the same vorticity $L$ could be obtained for different combinations 
of the applied field and the size of the square, e.g., $L = 6$ was found for $H = 60$~Oe, 
$a = 2$~$\mu$m and $H = 40$~Oe, $a = 2.5$~$\mu$m –- see images in Figs.~8(b) and 7(e), 
respectively. 
Sometimes two different vorticities were found in the same experiment for nominally 
identical squares, e.g., both $L = 9$ and $L = 10$ were found for $H = 35$~Oe and 
$a = 3.5$~$\mu$m –- see images in Figs.~7(g), (h), (i). 
The latter finding can be explained by slightly different shapes of individual 
squares or by an extra vortex captured during field cooling –- see Ref.~\cite{irina06} 
for a more detailed discussion, where the same effect was found for circular mesoscopic 
disks. 
Overall, the vorticity as a function of the applied field $H$ showed the same behaviour 
as that found earlier for circular disks \cite{irina06}, i.e., the square dots showed 
strong diamagnetic response for small vorticities $L < 10$ 
(also observed earlier in disks with a strong disorder \cite{irina07})
while for larger vorticities 
the extra demagnetisation per vortex saturated at $\delta \Phi / \Phi \approx 0.2$, 
in excellent agreement with earlier numerical studies \cite{baelus_prb02}. 

Most of the vortex configurations shown in Fig.~7 represent just one of several 
possible states for each vorticity (with the exception of images (h) and (i) which 
both correspond to $L = 10$). 
Indeed, for most vorticities we found more than one well-defined vortex configuration 
and some of these were found with almost the same probability, indicating that, 
in agreement with theory described above, vortices in mesoscopic squares form not 
only the ground, but also metastable states, and the energies of the latter are 
often very close to the energy of the ground state. 
This conclusion follows from our statistical analysis of all observed vortex 
configurations which resulted in histograms such as those shown in Fig.~8 for 
$L = 2$, 4, 5, and 6. 
For $L = 2$ and 4, the most frequently observed states agree with the ground states 
found theoretically (see Fig.~3(b), (d)) and the metastable states appear to have 
similar energies, as they are found with similar probabilities. 
As expected, both states for $L = 2$ and two of the states for $L = 4$ have vortices 
sitting along the symmetry axes of the square, with the diagonal axis being slightly 
preferable. 
The third state for $L = 4$ (on the right-hand side in Fig.~8(a)) is more unusual 
in that the vortices are sitting in the apexes of a rhombus that is slightly rotated 
with respect to the diagonal of the square. 
Although this particular state did not come out in the numerical simulations \cite{rhombus}, 
it was found with a high probability in experiment and, moreover, the rhombus-based 
vortex configurations were also found for larger vorticities both in experiment 
(see, e.g., Fig.~7(l) for $L = 13$) and theory (see rhombic inner shells for 
$L = 12$ and 16 in Figs.~3(l) and 4(b), respectively). 

For $L = 6$, one of the two most frequently observed states (also shown in Fig.~7(e)) 
corresponds exactly to the ground state found numerically (Fig.~3(f)) but the state 
found in experiment with the highest probability is the more symmetric two-shell 
configuration with the outer shell having the same pentagon shape as that found 
for $L = 5$. 
This $L = 6$ state can be viewed as a direct precursor of the two-shell states for 
$L = 7$ and 9, which were found as ground states both in theory (Fig.~3(g),(i)) 
and experiment (Fig.~7(f),(g)).  
For $L=5$, two possible states -- a two-shell configuration with one vortex in the 
center (1,4) and four vortices in the corners and a pentagon-like configuration (5) -- 
were found in experiment and in numerical simulations. 
However, numerical simulations found a slightly lower energy for the two-shell 
configuration 
(1,4) (see Fig.~5), while in experiment the pentagon-shaped configuration was found 
to appear more frequently. 
This discrepancy is unlikely to be related to the non-ideal character 
of the experimental squares: As we show below, neither the roughness of the 
boundaries, nor the presence of  some pinning in the experimental samples 
have any noticeable effect on the observed vortex configurations, due to 
strong confinement (see, e.g., Fig.~2). 
It is possible that, due to the very 
small difference in free energies between the two states
(which becomes practically negligible for samples with $a \gg \lambda$), 
the vortex 
configurations for $L=5$ are particularly sensitive to the exact sample 
size (in experiment the squares are almost 10 times larger than in the 
analysis of Fig.~5). 
The sensitivity of vortex configurations to sample 
size was studied in detail for circular disks (see Ref.~\onlinecite{slava_disk}) 
and was indeed 
found to affect the stability of some (but not all) vortex states. 
For higher vorticilties, $L = 7$ to 13, we found well defined two-shell configurations 
most of which correspond to the stable configurations found numerically. 
The outer shell in these configurations was either square (see Figs.~7(g)-(k) for 
$L = 9$ to 12), circular ($L = 7$, Fig.~7(f)) or rhombic ($L = 13$, Fig.~7(l)) with 
vortices of the inner shell either sitting along one of the symmetry axes of the square, 
as for $L = 2$, or forming a triangle, as for $L = 3$. 
For certain matching vorticities ($L = 9$ and 12), the observed two-shell configurations 
correspond to a square vortex lattice.

\begin{figure}
\begin{center}
\includegraphics[width=0.95\columnwidth]{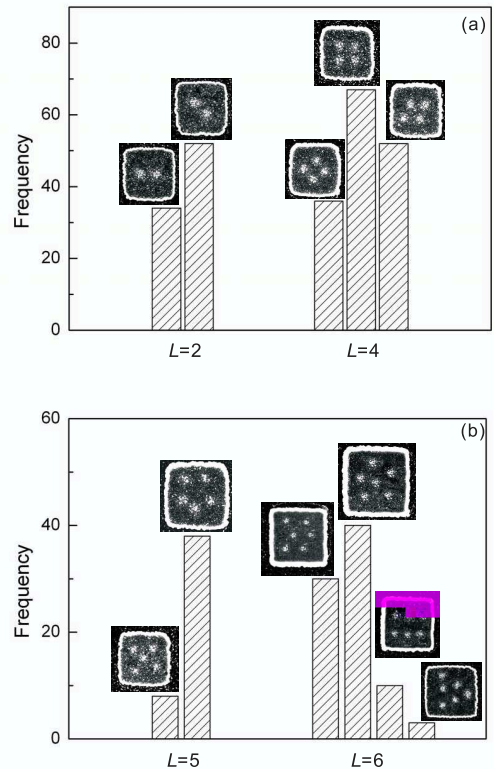}
\caption{
Histograms of different vortex states observed for vortcities $L = 2$, 4 
(for squares with $a=2\mu m$) (a) 
and $L = 5$ (for squares with $a=2\mu m$) and 6 (b) ($a=2\mu m$ and $a=2.5\mu m$). 
SEM images of the corresponding vortex configurations are shown as insets. 
}
\label{sqfig08}
\end{center}
\end{figure}

We note that the irregularities of the sample shape and uneven boundaries of some of our 
dots have, surprisingly, no discernable effect on the observed configurations of vortices
(i.e., the vortices form regular, symmetric patterns). 
For example, the dots in Figs.~7(j),(k) have especially rounded corners and very rough 
boundaries but the vortex configurations have square symmetries. 
Similarly, the same $L = 6$ state was found in dots with rounded corners, as in Fig.~7(e), 
and in almost perfect squares, as in the image shown in Fig.~8(b). 
Furthermore, we found that for a given value of $L$ the observed configurations did not 
depend on the sample size or the applied field, at least within the studied field range 
– see Fig.~9 for an example.

\begin{figure}
\begin{center}
\includegraphics[width=0.95\columnwidth]{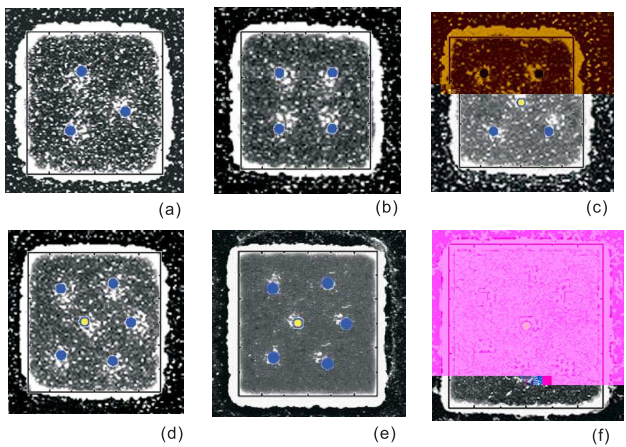}
\caption{
(Color online) 
Comparison of the experimentally observed positions of vortices within the square dots 
with those found numerically. 
Superimposed on the experimental images are vortex configurations shown in Figs.~3(c),(d),(e),(f),(g). 
Two experimental images for $L = 6$  ((d) and (e)) are superimposed on the same theoretical 
image (Fig.~3(f)), to demonstrate that the observed configurations did not depend on the 
sample size or the applied field (for image (d) $H = 40$~Oe, $a \approx 2.5$~$\mu$m, 
for image (e) $H = 60$~Oe, $a \approx 2$~$\mu$m). 
}
\label{sqfig09}
\end{center}
\end{figure}

Finally, we compared the experimentally observed positions of vortices within the square 
dots with those found numerically and found an excellent agreement, as demonstrated by Fig.~9. 
Here we show a superposition of theoretical images from Fig.~3 and experimental images for 
the same vortex configurations. 
Two of the images (Figs.~9(d),(e)) compare the same theoretical configuration with experimental 
images obtained on dots of different sizes in different applied fields ($H = 40$~Oe, $a = 2.5$~$\mu$m 
and $H = 60$~Oe and $a = 2$~$\mu$m, respectively) illustrating the point made above that the 
vortex configurations do not depend on the sample size and/or applied field. 

Overall, despite the inevitable presence of some disorder in our samples, 
which was not taken into account in the calculations, 
there is a very good agreement between the observed vortex configurations 
and the calculated vortex patterns. 
The main features of the vortex states revealed by experiment is formation of vortex shells 
with predominantly square symmetry for vorticities $L \geq 7$ and vortex patterns following 
the main symmetry axes of the square for small vorticities $L \leq 4$. 
The two intermediate vorticities $L = 5$ and 6 appear to be a special case: Here the mismatch 
between the square shape of the dot and the natural symmetry of the vortex lattice is more 
difficult to accommodate and the preferred vortex configurations turned out to be the 
pentagon-shaped shell for $L=5$ and three different patterns for $L=6$, none of which has 
the four-fold symmetry of the square.

\section{Conclusions}

We performed a systematic study of vortex configurations 
in mesoscopic superconducting squares and compared the results 
with vortex patterns observed experimentally in $\mu$m-sized Nb 
squares using the Bitter decoration technique. 

In the theoretical analysis we relied upon the analytical solution 
of the London equation in mesoscopic squares by using the Green's 
function method and the image technique. The stable vortex 
configurations were calculated using the technique of 
molecular-dynamics simulations simulating the stimulated annealing 
process in experiments. 

We revealed the filling rules for squares with growing number of 
vortices $L$ when gradually increasing the applied magnetic field. 
In particular, we found that for small $L$ vortices tend to form 
patterns that are commensurate with the symmetry of the square 
boundaries of the sample. The filling of ``shells'' (similar to 
mesoscopic disks) occurs by periodic filling of the outermost and 
internal shells. With increasing vorticity, the outermost shell is 
filled until it is complete (i.e., the number of vortices in it
becomes $4N$, where $N$ is an integer, i.e., commensurate with the
square boundary). Then vortices fill internal shells untill the
number of vortices becomes large enough to create the outermost
shell with $4(N+1)$ vortices. Again, after that vortices fill
internal shells. With increasing vorticity, the shell structure
becomes less pronounced, and for large enough $L$ the vortex
patterns in squares becomes a traingular lattice distorted near
the boundaries.

\section{Acknowledgments}

We thank Mauro M. Doria for useful discussions. 
This work was supported by the Flemish Science Foundation (FWO-Vl), 
the Interuniversity Attraction Poles (IAP) Programme $-$ 
Belgian State $-$ Belgian Science Policy, the ``Odysseus'' program 
of the Flemish Government and FWO-Vl, and EPSRC (UK). 
V.R.M. is funded by the EU Marie Curie project, 
Contract No. MIF1-CT-2006-040816.

\end{document}